\documentclass{article}

\usepackage{amsfonts, amsmath, amssymb, amsbsy}
\usepackage{graphicx}

\begin{document}

\begin{center}
\textbf{\LARGE AGB stars and the plate archives heritage.}

{presented at ASTROPLATES 2014, Prague 18-21 March 2014}
\end{center}

\begin{center}
\textbf{Roberto Nesci $^1$, Silvia Gaudenzi$^1$, Corinne Rossi$^2$, Camilla Pezzotti$^2$, Kamo Gigoyan$^3$, Nicolas Mauron$^4$
}
\end{center}

\begin{center}
{\it
\noindent 
$^1$INAF-IAPS, Roma, Italia \\
$^2$Universita' La Sapienza, Roma, Italia \\
$^3$Byurakan Observatory, Byurakan, Armenia \\
$^4$CNRS \& Universit\'e Montpellier II, Montpellier, France \\
 }
\end{center}

\begin{abstract}
We report on the characterization of a number of AGB candidate stars identified with objective-prism plates of the Byurakan Observatory. 
Digitized photographic sky survey plates and recent CCD photometry have been used to improve the selection and distinguish variable and non-variable stars. Some comparisons among published catalog magnitudes are also made.
Slit spectroscopy from the Asiago and Loiano Observatories allowed a firm spectral classification, separating C-Type, N-Type and normal M giants.
Color-color plots using WISE, AKARI and 2MASS J-band data allow an efficient discrimination of spectral types, which can be used for the definition of larger statistical samples.

\end{abstract}

\section{Introduction}
The large Schmidt telescope of the Byurakan Observatory (102/132/213 cm) with the 1.5 degree prism and 103aF emulsion made a successful survey of the northern sky in the 1970's. On these plates late type stars are easily recognized by their strong, nearly point-like, red emission while the blue part of the spectrum is very faint or even absent. 
This allows to build sample of AGB stars candidates, and/or to check the nature of infrared (IR) objects detected by space-based IR instruments like WISE, IRAS, AKARI or of ground-based surveys like 2MASS.

The first Byurakan Survey plates and automatically extracted spectra are freely accessible at http://byurakan.phys.uniroma1.it/index.php (Mickaelian et. al. 2007).

The AGB stars, due to their large luminosity, can be traced up to the limit of our Milky Way halo and therefore serve as tracers of its gravitational field and of dark matter distribution (see e.g. Mauron et al. 2013 and references therein).
In this paper we report on an on-going survey program of AGB candidates in the northern hemisphere, 
aimed at obtaining a better spectral classification and exploring the possibilities of a classification from recent photometric infrared data from satellite surveys. Candidate stars were taken from the "Revised and Updated Catalogue of FBS Late-Type Stars"( Gigoyan and Mickaelian, 2012).

\section{Spectroscopic data}
Each star has been spectroscopically classified using the BFOSC instrument at the Loiano 1.5m telescope, and the AFOSC instrument at the Asiago 1.8m telescope, with resolution of 12 \AA and in many cases of 6 \AA \footnote{Partially based on observations taken at the Asiago and Loiano Observatories}. We are therefore able to establish a firm spectral subtype, distinguish Early- and Late- types of Carbon stars, normal Oxigen stars, and discriminate giants from dwarfs. Actually very few dwarfs were found.

In Fig. 1 left panel, the appearance of a Carbon star (FBS2213+421) on a Byurakan plate is shown.

\begin{figure}
\includegraphics[width=15cm]{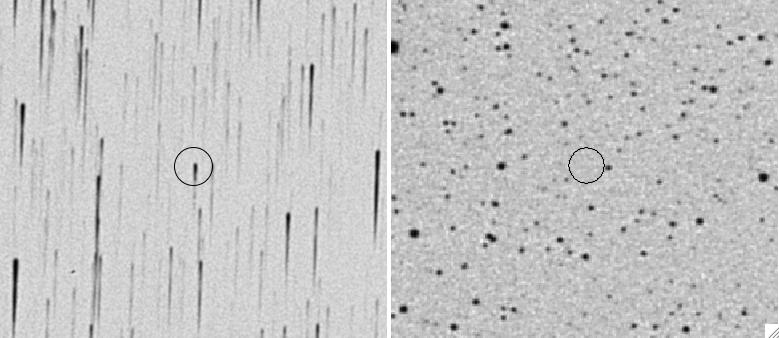}
\caption{Left panel: Byurakan objective-prism plate n.936 taken in 1973 showing the C-type star FBS2213+421 (V 381 Lac) at B=15.6.
Right panel: Asiago plate n.6394, taken in 1967 showing the field of the star, which was below the plate limit B=17.7 .}
\end{figure}

\section{Variability}
Among the stars of our sample, several are variable. As an example, from the  data available in literature, V381 Lac was very  bright at the epoch of POSS1 (1952); invisible in the red plates of POSS2 (1989);  bright  between 1999 and 2000, from NSVS;  undetected  by the SDSS in 2006.  From our archives we see that the star shows a well exposed spectrum in the Byurakan plate (1973) while it is not detectable in the Asiago  direct image of 1967 shown in right panel of Fig 1.

We started checking the variability of our stars using the USNO-B1 catalog, based on the POSS I and POSS II surveys, in the B and R bands. The photometric accuracy of this catalog is not high, especially for stars partially saturated: form our experience a difference of at least 0.5 magnitudes is required to be sure of the star variability.

A few other digitized plates taken with the Palomar 120/180/307 cm Schmidt, besides those used for the Digital Sky Survey, can be retrieved from the Space Telescope Science Institute website 
(http://archive.stsci.edu/cgi-bin/dss-plate-finder/).  To derive magnitudes also from these plates,
we re-made the aperture photometry of our stars with IRAF/apphot on all the POSS plates, using some nearby (radius 4 arcmin) reference stars taken from the GSC2.3.2 catalog: for consistency, the same sequences were used also for the CCD photometry of our images from the Asiago and Loiano telescopes.

It is useful to have an estimate of the photometric reliability of the most used all sky catalogs.
In our search we tested the USNO-A2, the USNO-B1 and the GSC2.3.2. The new APASS (UCAC 4), still not covering the whole sky, may become a valid alternative in the near future, given that it is not based on photographic photometry but is entirely based on CCD images.

The photometric calibration of the DSS plates for the aforementioned catalogs was made on a plate-by-plate basis. Differences between catalogs are therefore expected to vary somehow from plate to plate:
we report here the relations between catalogs for the field of V 381 Lac.

\noindent
R(USNO-A2)=B(GSC)*0.86+2.63 (rms=0.27)\\
B(USNO-A2)=B(GSC)*0.80+2.71  (rms=0.43).

The correlation is very good, but for the R magnitudes there is a small deviation from a straight line for the bright (R$<$13) stars. Indeed using only GSC2 stars with R$<$15 (35) the slopes become more near to the ideal slope 1.

\noindent
R(USNO-A2)=R(GSC)*0.99+1.77 (rms=0.26)\\
B(USNO-A2)=B(GSC)*0.83+2.34 (rms=0.36)

\medskip
The comparison with USNO-B1 is the following:

\noindent
R1(usno)=R(gsc)*1.03-0.49 (rms=0.23)\\
R2(usno)=R(gsc)*1.02-0.12 (rms=0.27)\\
B1(usno)=B(gsc)*0.99+0.08 (rms=0.28)\\
B2(usno)=B(gsc)*0.96+0.94  (rms=0.42)\\

Comparison between the USNO-A2 and USNO-B1 is good, with systematic deviation for bright stars (B$<$15).

\noindent
B1(usno)=B(a2)*1.23-3.18 (rms=0.33) \\
B2(usno)=B(a2)*1.16-1.90 (rms=0.38) \\

\medskip

Further sources of photometric data are the recent sky surveys for NEO or GRB, like the Catalina Sky Survey at Caltech (Drake et al. 2009; http://nesssi.cacr.caltech.edu/DataRelease/) or the NSVS (Wozniack et al. 2004; http://skydot.lanl.gov/nsvs/nsvs.php). 

We checked the zero-point consistency of the magnitude scales of these Surveys with the GSC2.3.2 catalog, using some stars which appeared to be nearly constant in the Surveys.
It is worth to remember that the NSVS data were obtained with an unfiltered CCD,
so that the quantum efficiency of the sensor makes the effective band most comparable to
the Johnson R band (Wozniak et al. (2004)), or better a mix of V and R colors, which is
a function of the spectral type of the star.
To inter-calibrate the NSVS and our magnitudes we used the stars in our sample with
a very stable NSVS light curve.  We found that a reliable calibration would require
a bigger set of non variable stars in the M0-M8 spectral types range to define a better
color correction. In fact, we verified that for M8 stars, which emit most the photons in
the IR tail of the unfiltered detector, the ROTSE instrumental magnitude are generally
brighter than for M1-type stars of similar R magnitude. In any case, even with this caveat, our
data have been useful to confirm the variability/stability of the stars of our sample.

\section{Color selection}
As supplementary diagnostic for stellar classification we used IR photometry by applying the color--magnitude and color--color diagrams to data downloaded from several ground based and satellite surveys. 

 \begin{figure}
 \includegraphics[width=14cm]{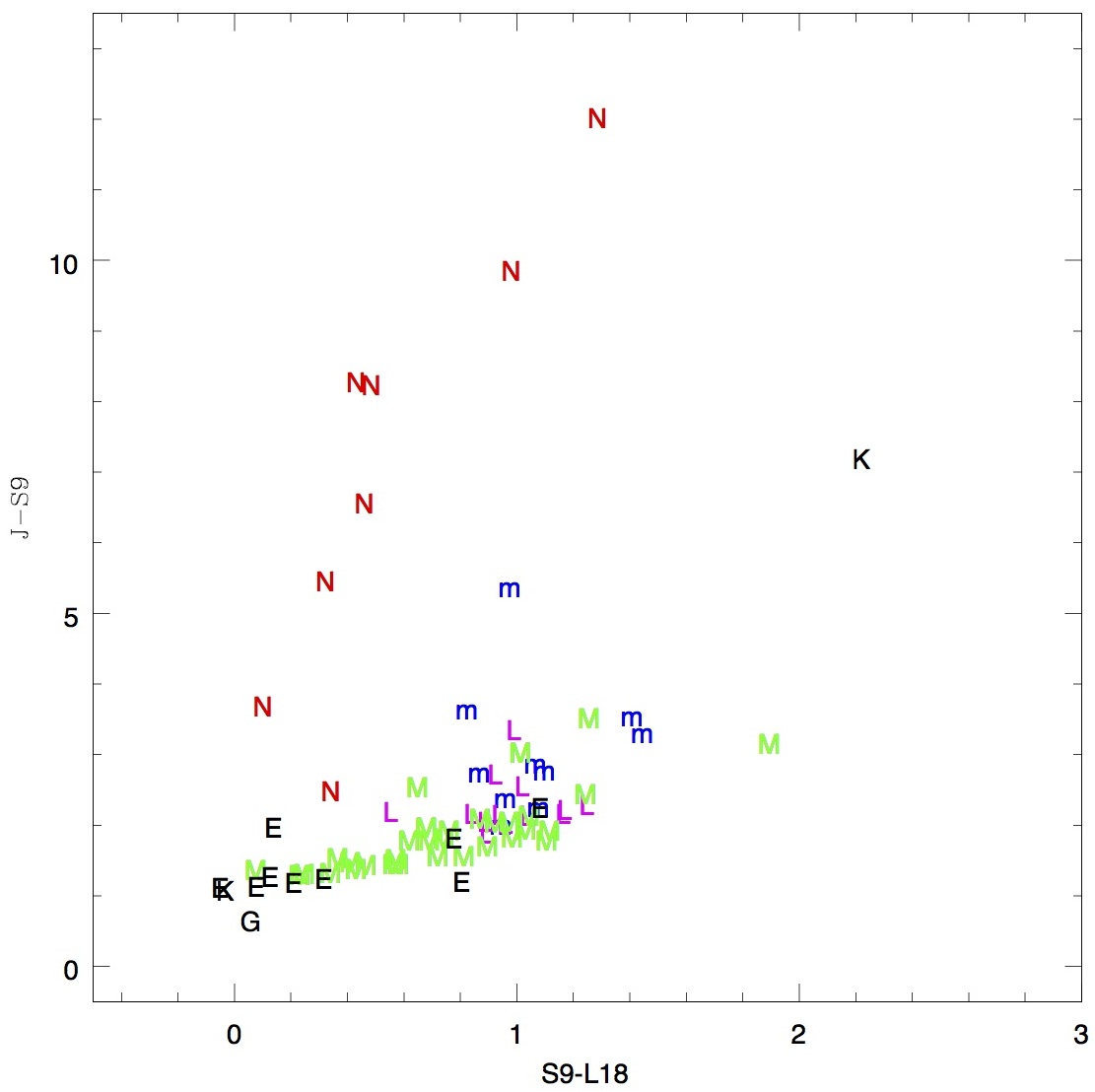}
 \caption{ AKARI-Jband color-color plot of the sample stars. L=M5 and later; E=Earlier than M5; N=N-type stars; C= CH stars; M=generic M-type stars; m=Mira variable; K and F mark peculiar stars. Abscissa AKARI S9-L18 magnitude difference; ordinate J-AKARI L18 magnitude difference.}
 \label{cartin2}
\end{figure}

We checked among the different color-color plots which can be made using the AKARI, WISE, 2MASS fluxes of our sample stars if there are good photometric discriminants between Carbon rich and Oxygen rich objects. Extensive work on these  tools has been made by Ishihara et al. (2011) and Xun Tu \& Zhong-Xian Wang (2013). Here we present preliminary results using different color choices.
Two examples are reported in Fig. 2 and 3. Figure 2 shows that the J-L18 color index, coupled with the AKARI S9-L18 index provides a very good selection between Carbon stars, Nitrogen stars and normal M-type stars. 

Several useful diagrams can be obtained from the combination of the WISE colors: in this paper we selected the 3.4-12 $\mu$m (w1-w3) {\it vs} 12-22$\mu$m  (w3-w4). This diagram is the best among the WISE ones to discriminate the different  types of stars, establishing  a well defined sequence of locations for variability class and separating all the  Carbon  from the M stars. Fig. 3 shows how  our sample stars are located in this color--color plot. 

 \begin{figure}
 \includegraphics[width=14cm]{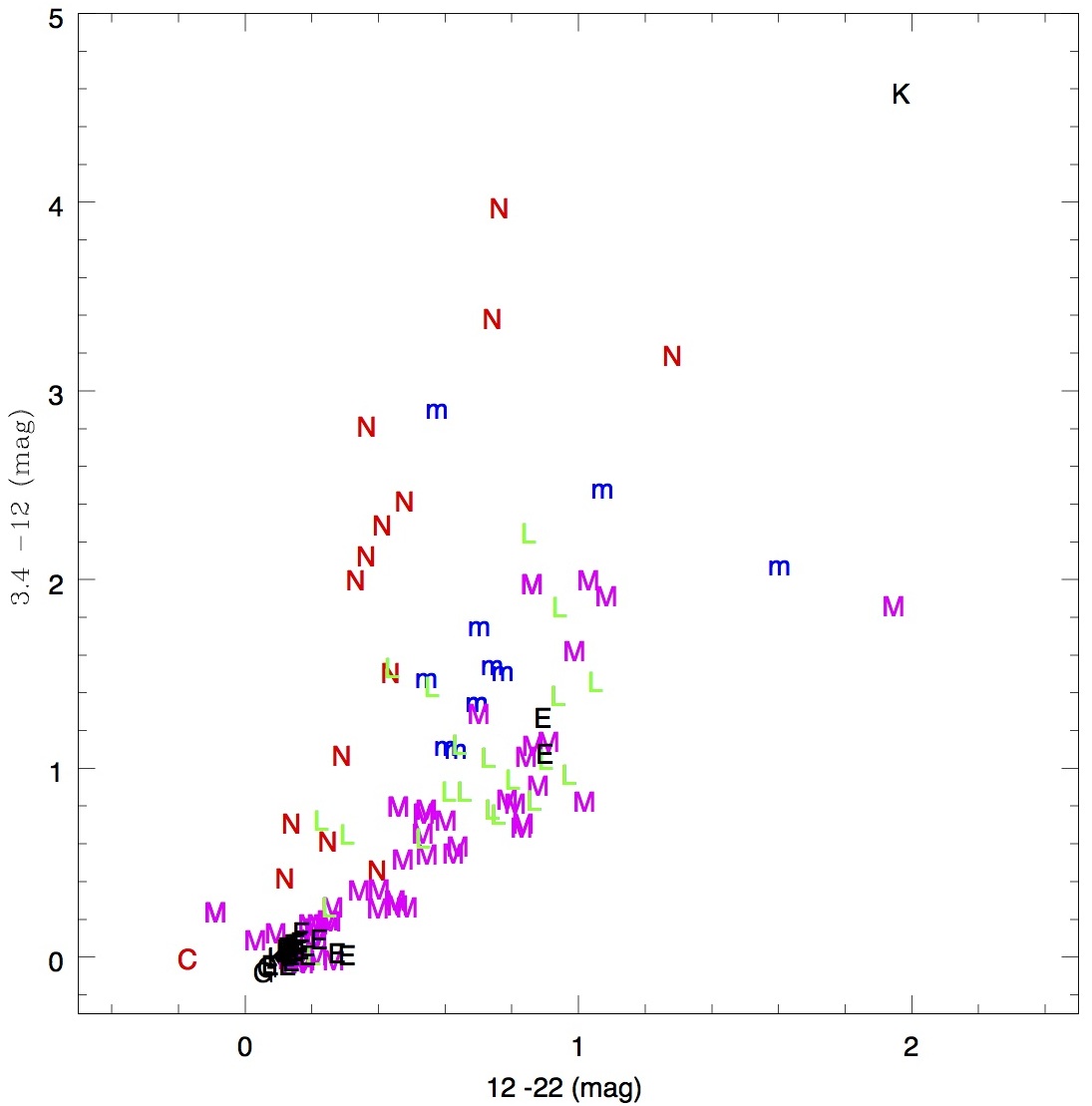}
 \caption{ WISE color-color plot; abscissa W3-W4, ordinate W1-W3. Star types as in Fig. 2
 }
 \label{cartin3}
\end{figure}

\medskip
 
As a further check we looked also at the IRAS catalog to locate our sources in the classical [25]-[60] vs [12]-[25] color--color plot. The agreement with the work by van der Veen and  Habing (1988)  is poorer than expected.  The most probable reason of some strong discrepancies might be the large uncertainties of the fluxes reported in the IRAS catalog for several of our stars, suggesting to use great caution in the application of this plot.
  
\section{Conclusions}
Photographic archives can still contribute to high quality science, provided that they can be easily accessed from the web. A fast consultation of images, even if of not high quality, may indeed help the researcher to have a quick response and decide if further investigation is worth to be made.

The technology of plate scanners is rapidly evolving, as well as the availability of very massive electronic storage at low price, so the task  of digitization is easier than just 10 years ago. A good storage of the original photographic material is therefore important, because in few years better scanners may be available allowing more accurate data retrieval.

Putting on line at least the logbooks of the plate archives is mandatory making the old photographic data 
accessible and  efficiently used. 

\vskip 3cm

{\bf Acknowledgements}
This publication makes use of data products from the Wide-field Infrared Survey Explorer, which is a joint project of the University of California, Los Angeles, and the Jet Propulsion Laboratory/California Institute of Technology, funded by the National Aeronautics and Space Administration.

We thank the Directorate of the Loiano and Asiago Observatories for telescope time allocation

\end{document}